\documentclass{article}%
\usepackage{amsfonts}
\usepackage{amssymb}
\usepackage{amsmath}
\usepackage{graphicx}%
\providecommand{\U}[1]{\protect\rule{.1in}{.1in}}

\begin{document}

\title{From Abstract Entities in Mathematics to \\Superposition States in Quantum Mechanics}
\author{David Ellerman\\University of California at Riverside}
\maketitle
\tableofcontents

\begin{abstract}
\noindent Given an equivalence relation $\thicksim$ on a set $U$, there are
two abstract notions of an \textit{element} of the quotient set $U/\thicksim$.
The $\#1$ abstract notion is a set $S=\left[  u\right]  $ of equivalent
elements of $U$ (an equivalence class); the $\#2$ notion is an abstract entity
$u_{S}$ that is definite on what is common to the elements of the equivalence
class $S$ but is otherwise indefinite on the differences between those
elements. For instance, the $\#1$ interpretation of a homotopy type is an
equivalence class of homotopic spaces, but the $\#2$ interpretation, e.g., as
developed in homotopy type theory, is an abstract space (without points) that
has the properties that are in common to the spaces in the equivalence class
but is otherwise indefinite. In philosophy, the $\#2$ abstract entities might
be called \textit{paradigm-universals}, e.g., `\textit{the} white thing' as
opposed to the $\#1$ abstract notion of "the set of white things" (out of some
given collection $U$).

The paper shows how this $\#2$ notion of a paradigm may be mathematically
modeled using incidence matrices in Boolean logic and density matrices in
probability theory. Then we cross the bridge to the density matrix treatment
of the indefinite superposition states in quantum mechanics (QM). This
connection between the $\#2$ abstracts in mathematics and ontic indefinite
states in QM elucidates Abner Shimony's literal or objective indefiniteness
interpretation of QM.

\end{abstract}

\section{Introduction}

The purpose of this paper is to illuminate the late Abner Shimony's
objectively indefinite or `Literal' interpretation of quantum mechanics based
on seeing the superposition states as being objectively indefinite.

\begin{quotation}
\noindent From these two basic ideas alone -- indefiniteness and the
superposition principle -- it should be clear already that quantum mechanics
conflicts sharply with common sense. If the quantum state of a system is a
complete description of the system, then a quantity that has an indefinite
value in that quantum state is objectively indefinite; its value is not merely
unknown by the scientist who seeks to describe the system. \cite[p.
47]{shim:reality}
\end{quotation}

In addition to the Shimony's phrase "objective indefiniteness," other
philosophers of physics have used similar phrases for these indefinite states:

\begin{itemize}
\item Peter Mittelstaedt's "incompletely determined" quantum states with
"objective indeterminateness" \cite{mitt:kant};

\item Paul Feyerabend's "inherent indefiniteness" \cite{feyerabend:micro};

\item Allen Stairs' "value indefiniteness" and "disjunctive facts"
\cite{stairs:disjfacts};

\item Steven French and Decio Krause's "ontic vagueness"
\cite{french:onticvagueness}; or

\item E. J. Lowe's "vague identity" and "indeterminacy" that is "ontic"
\cite{lowe:vagueid}.
\end{itemize}

\noindent But how can we understand the notion of an "ontic indefinite state"?

\section{Two Versions of Abstraction}

The claim is that we already have the notion of an indefinite state in the
mathematical notion of an entity that abstracts as definite what is common to
the distinct elements of a set $S$ and rendering their differences as indefinite.

Given an equivalence relation on a set $U$ such as "having the same color" and
if $u\thicksim u^{\prime}$ were white, then there are two notions of abstraction:

\begin{enumerate}
\item the $\#1$ version of the abstraction operation takes equivalent entities
$u\thicksim u^{\prime}$ to the equivalence class $\left[  u\right]  =\left[
u^{\prime}\right]  $ of all white entities (in some universe $U$), and;

\item the $\#2$ version of the abstraction operation takes all the equivalent
entities $u\thicksim u^{\prime}$ to the abstract entity "\textit{the} white
entity" that is definite on what is common in the set of all particular white
things but is indefinite on how they differ (e.g., on all the other properties
that distinguish them).
\end{enumerate}

For instance, there are two notions of an `element' of a quotient set or a
quotient group (or any other quotient object in algebra):

\begin{enumerate}
\item a quotient group element as an equivalence class or coset; or

\item a quotient group element as an abstract entity representing what is
common to the equivalence class.
\end{enumerate}

Given \textit{any }property $S\left(  u\right)  $ defined on the elements of
$U$, two abstract objects can be defined:%

\begin{center}
\includegraphics[
height=1.5152in,
width=4.6639in
]%
{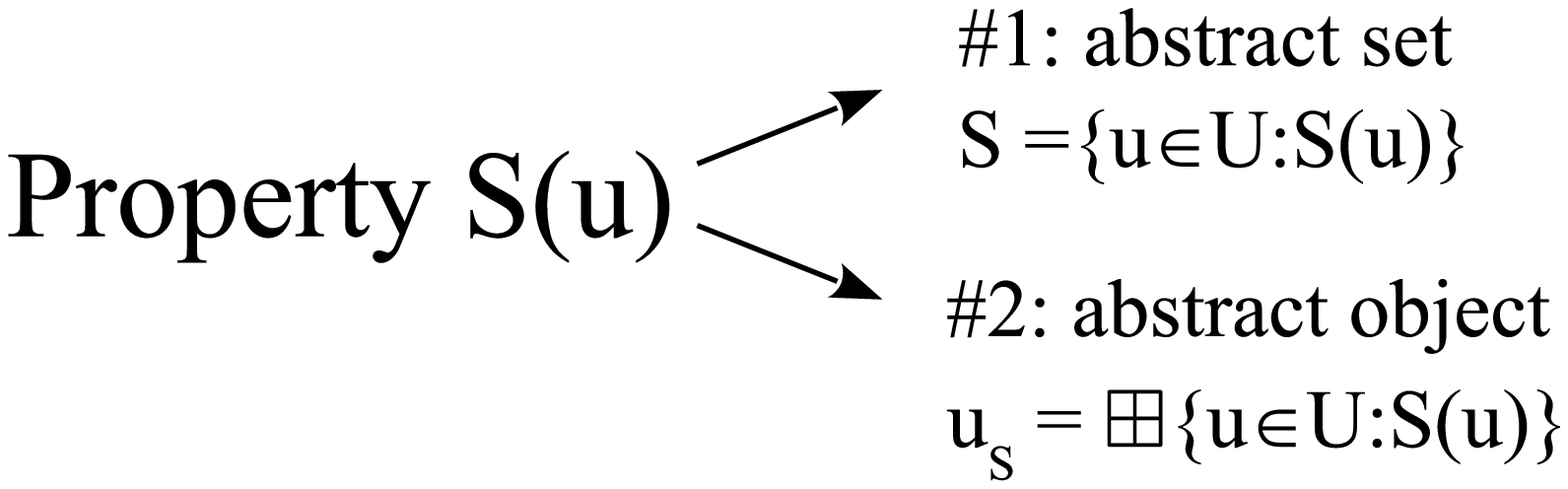}%
\end{center}

\begin{center}
Figure 1: A property determines two types of abstract objects.
\end{center}

Intuitively the $\#2$ abstract object $u_{S}$ is `\textit{the} paradigm
$S$-entity' (the blob-sum $\boxplus$ is defined below) which is definite on
the $S\left(  u\right)  $ property and indefinite on (i.e., blobs out) the
differences between all the $u\in U$ such that $S\left(  u\right)  $.

\section{An Example Starting with Attributes}

Consider three predicates (binary attributes) $P\left(  x\right)  $, $Q\left(
x\right)  $, and $R\left(  x\right)  $ which could distinguish at most
$2^{3}=8$ definite-particular entities: $u_{1},...,u_{8}$ called
\textit{eigen-elements} and which can be presented in a table like a truth table:

\begin{center}%
\begin{tabular}
[c]{|c|c|c|c|}\hline
$P\left(  x\right)  $ & $Q\left(  x\right)  $ & $R\left(  x\right)  $ &
$u$\\\hline\hline
$1$ & $1$ & $1$ & $u_{1}$\\\hline
$1$ & $1$ & $0$ & $u_{2}$\\\hline
$1$ & $0$ & $1$ & $u_{3}$\\\hline
$1$ & $0$ & $0$ & $u_{4}$\\\hline
$0$ & $1$ & $1$ & $u_{5}$\\\hline
$0$ & $1$ & $0$ & $u_{6}$\\\hline
$0$ & $0$ & $1$ & $u_{7}$\\\hline
$0$ & $0$ & $0$ & $u_{8}$\\\hline
\end{tabular}

Table 1: Eight entities specified by 3 properties.
\end{center}

The general rule is if $f,g,h:U\rightarrow%
\mathbb{R}
$ are numerical attributes with the number of distinct values as $n_{f}$,
$n_{g}$, and $n_{h}$ respectively, then those attributes could distinguish or
classify $n_{f}\times n_{g}\times n_{h}$ distinct subsets of $U$. If the join
of the inverse-image partitions is the discrete partition, i.e.,$\left\{
f^{-1}\right\}  \vee\left\{  g^{-1}\right\}  \vee\left\{  h^{-1}\right\}
=\mathbf{1}_{U}$ \cite{ell:partitions}, then $\left\{  f,g,h\right\}  $ is a
\textit{complete set of attributes} since they can distinguish or classify the
eigen-elements of $U$. Then we can distinguish the elements of $U$ by their
triple of values, i.e., $\left\vert f\left(  u_{j}\right)  ,g\left(
u_{j}\right)  ,h\left(  u_{j}\right)  \right\rangle $ uniquely determines
$u_{j}\in U$.

In the example, any subset $S\subseteq U=\left\{  u_{1},...,u_{8}\right\}  $
is characterized by a property $S\left(  x\right)  $, the \textit{disjunctive
normal form property}, common to all and only the elements of $S$. If
$S=\left\{  u_{1},u_{4},u_{7}\right\}  $, then the DNF property is:

\begin{center}
$S\left(  x\right)  =\left[  P\left(  x\right)  \wedge Q\left(  x\right)
\wedge R\left(  x\right)  \right]  \vee\left[  P\left(  x\right)  \wedge\lnot
Q\left(  x\right)  \wedge\lnot R\left(  x\right)  \right]  \vee\left[  \lnot
P\left(  x\right)  \wedge\lnot Q\left(  x\right)  \wedge R\left(  x\right)
\right]  $.
\end{center}

But what are the $\#1$ and $\#2$ abstract entities?

\begin{enumerate}
\item[1.] The $\#1$ abstract entity is the set
\end{enumerate}

\begin{center}
$S=\left\{  u_{i}\in U|S\left(  u_{i}\right)  \right\}  =\left\{  u_{1}%
,u_{4},u_{7}\right\}  $
\end{center}

of all the distinct $S\left(  x\right)  $-entities; and

\begin{enumerate}
\item[2.] The $\#2$ abstract entity is \textit{the paradigm-universal}
$S\left(  x\right)  $-entity symbolized
\end{enumerate}

\begin{center}
$u_{S}=u_{1}\boxplus u_{4}\boxplus u_{7}=\boxplus\left\{  u_{i}\in U|S\left(
u_{i}\right)  \right\}  $

The `superposition' or `blob-sum' of $u_{1}$, $u_{4}$, and $u_{7}$.
\end{center}

\noindent that is \textit{definite on the DNF property }$S\left(  x\right)
$\textit{\ }but indefinite on what distinguishes the different $S\left(
x\right)  $-entities. Thus $S\left(  u_{S}\right)  $ holds but none of the
disjuncts hold since that would make $u_{S}$ equal to $u_{1}$, $u_{4}$, or
$u_{7}$. Hence $S\left(  u_{S}\right)  $ is a `disjunctive fact' in the sense
of Allen Stairs \cite{stairs:disjfacts}.

\section{Some Philosophical Concerns}

It is best to think of $S$ as the set of \textit{definite particular}
$S\left(  x\right)  $-entities in some universe $U$, while $u_{S}$ is the
\textit{indefinite paradigm-universal }$S\left(  x\right)  $-entity is the
`superposition' $u_{S}=\boxplus\{u_{i}\in U|S\left(  u_{i}\right)  \}$ that
is, in general, "one \textit{over} the many." Only when $S=\left\{
u_{j}\right\}  $ is a singleton does the definite description `\textit{the}
$S$-entity' refer to an element of $U$, i.e., $u_{\left\{  u_{j}\right\}
}=u_{j}$.

Making the "one" $u_{S}=\boxplus\{u_{i}\in U|S\left(  u_{i}\right)  \}$
\textit{over} the many, i.e., more abstract than the $u_{i}\in U$ (for
$\left\vert S\right\vert >1$) avoids the paradoxes just as the iterative
notion of set does in ordinary set theory, i.e., for $\#1$ type of
abstractions. Otherwise, if we ignore the given set $U$, then we can recreate
Russell's Paradox for $R\left(  u_{S}\right)  \equiv\lnot S\left(
u_{S}\right)  $ so:

\begin{center}
$u_{R}=\boxplus\left\{  u_{S}|\lnot S\left(  u_{S}\right)  \right\}  $ and
thus $R\left(  u_{R}\right)  $ implies $\lnot R\left(  u_{R}\right)  $, and
$\lnot R\left(  u_{R}\right)  $ implies $R\left(  u_{R}\right)  $.
\end{center}

\noindent But if we define $u_{R}=\boxplus\left\{  u_{S}\in U|\lnot S\left(
u_{S}\right)  \right\}  $, then assuming $u_{R}\in U$ leads to the
contradiction so $u_{R}\notin U$.

The paradigm-universal $u_{S}$ is \textit{not} universal `$S$-ness'. Where
$S\left(  x\right)  $ is being white, then $u_{white}=$ `\textit{the} white
thing`, not `whiteness'. This distinction goes back to Plato:

\begin{quotation}
\noindent But Plato also used language which suggests not only that the Forms
exist separately ($\chi\omega\rho\iota\sigma\tau\alpha$) from all the
particulars, but also that each Form is a peculiarly accurate or good
particular of its own kind, i.e., the standard particular of the kind in
question or the model ($\pi\alpha\rho\alpha\delta\varepsilon\iota\gamma
\mu\alpha$) to which other particulars approximate. \cite[p. 19]%
{kneales:logic}
\end{quotation}

\noindent Some have considered interpreting the Form as \textit{paradeigma} as
an error.

\begin{quotation}
\noindent For general characters are not characterized by themselves: humanity
is not human. The mistake is encouraged by the fact that in Greek the same
phrase may signify both the concrete and the abstract, e.g. $\lambda
\varepsilon\upsilon\kappa o\nu$ (literally "the white") both "the white thing"
and "whiteness", so that it is doubtful whether $\alpha\upsilon\tau o$ $\tau
o$ $\lambda\varepsilon\upsilon\kappa o\nu$ (literally "the white itself")
means "the superlatively white thing" or "whiteness in abstraction". \cite[pp.
19-20]{kneales:logic}
\end{quotation}

Thus for the abstract property $W\left(  u\right)  $ "whiteness", we have:

\begin{enumerate}
\item the $\#1$ abstraction is \textit{the set of white things} $W=\left\{
u\in U:W\left(  u\right)  \right\}  $, and;

\item the $\#2$ abstraction `\textit{the} white thing' $u_{W}$.
\end{enumerate}

\section{Relations Between \#1 and \#2 Universals}

For properties $S()$ defined on $U$, there is a 1-1 correspondence between the
$\#1$ and $\#2$ universals:

\begin{center}
$\cup\left\{  \left\{  u\right\}  |u\in U\&S\left(  u\right)  \right\}  =S$
$\longleftrightarrow$ $u_{S}=\boxplus\left\{  u_{\left\{  u\right\}  }|u\in
U\&S\left(  u\right)  \right\}  $.
\end{center}

In each case, we may extend the definition of the property to the two
universals. For $T()$ another property defined on $U$:

\begin{center}
$S\left(  T\right)  $ iff $\left(  \forall u\in U\right)  \left(  T\left(
u\right)  \Rightarrow S\left(  u\right)  \right)  $ iff $S\left(
u_{T}\right)  $.
\end{center}

In terms of the $\#1$ universals, $S\left(  S\right)  $ holds by definition
and: $S\left(  T\right)  $ iff $T\subseteq S$, and similarly $S\left(
u_{S}\right)  $ always holds. But what is the $\#2$ universals equivalent of
$T\subseteq S$? Intuitively $u_{S}$ is `\textit{the} $S$-thing' that is
definite on having the $S$-property but is otherwise indefinite on the
differences between the members of $S$. If we make more properties definite,
then in terms of subsets, that will in general cut down to a subset
$T\subseteq S$, so $u_{T}$ would inherit the paradigmatic property holding on
the superset $S$, i.e., $S\left(  u_{T}\right)  $.

This "process" to changing to a more definite universal $u_{S}\rightsquigarrow
u_{T}$ for $T\subseteq S$ will be called \textit{projection} and symbolized:

\begin{center}
$u_{T}\lhd u_{S}$ (or $u_{S}\rhd u_{T}$)

$u_{T}$ is a "sharpening" or more definite version of $u_{S}$.%

\begin{tabular}
[c]{|c|c|c|}\hline
$S()$ defined on $U$ & $\#1$ abstraction & $\#2$ abstraction\\\hline\hline
Universals for $S()$ & $S=\cup\left\{  \left\{  u\right\}  |u\in U\&S\left(
u\right)  \right\}  $ & $u_{S}=\boxplus\left\{  u_{\left\{  u\right\}  }|u\in
U\&S\left(  u\right)  \right\}  $\\\hline
$T()$ defined on $U$ & $S\left(  T\right)  $ iff $T\subseteq S$ & iff
$S\left(  u_{T}\right)  $ iff $u_{T}\lhd u_{S}$\\\hline
\end{tabular}

Table 2: Equivalents between $\#1$ and $\#2$ universals
\end{center}

In the language of Plato, the projection relation $\lhd$ is the relation of
"participation" ($\mu\varepsilon\theta\varepsilon\xi\iota\varsigma$ or
\textit{methexis}). As Plato would say, $u_{T}$ has the property $S()$ iff it
participates in `\textit{the} $S$-thing', i.e., $S\left(  u_{T}\right)  $ iff
$u_{T}\lhd u_{S}$.

Thus there are two theories of abstract objects:

\begin{enumerate}
\item Set theory is the theory of $\#1$ abstract objects, the sets $S$, where
(taking $\in$ as the participation relation), sets are \textit{never}
self-participating, i.e., $S\notin S$;

\item There is a second theory about the $\#2$ abstract entities, the
paradigms $u_{S}$, which are \textit{always} self-participating, i.e.,
$u_{S}\lhd u_{S}$.
\end{enumerate}

Like sets $S$, the $\#2$ abstract entities $u_{S}$, the
\textit{paradigm-universals}, are routinely used in mathematics.

\section{Examples of Abstract Paradigms in Mathematics}

There is an equivalence relation $A\simeq B$ between topological spaces which
is realized by a continuous map $f:A\rightarrow B$ such that there is an
inverse $g:B\rightarrow A$ so the $fg:B\rightarrow B$ is homotopic to $1_{B}$
(i.e., can be continuously deformed in $1_{B}$) and $gf$ is homotopic to
$1_{A}$. Classically "Homotopy types are the equivalence classes of spaces"
\cite{baues:ht} under this equivalence relation. That is the $\#1$ type of abstraction.

But the interpretation offered in homotopy type theory is expanding identity
to "coincide with the (unchanged) notion of equivalence" \cite[p. 5]{ufp:hott}
so it would refer to the $\#2$ homotopy type, i.e., `\textit{the} homotopy
type' that captures the mathematical properties shared by all spaces in an
equivalence class of homotopic spaces (wiping out the differences). Note that
`\textit{the} homotopy type' is not one of the classical topological spaces
(with points etc.) in the $\#1$ equivalence class of homotopic spaces.

\begin{quotation}
\noindent While classical homotopy theory is analytic (spaces and paths are
made of points), homotopy type theory is synthetic: points, paths, and paths
between paths are basic, indivisible, primitive notions. \cite[p.
59]{ufp:hott}
\end{quotation}

\noindent Homotopy type theory systematically develops a theory of the $\#2$
type of abstractions that grows out of homotopy theory and type theory in a
new foundational theory.

\begin{quotation}
\noindent From the logical point of view, however, it is a radically new idea:
it says that isomorphic things can be identified! Mathematicians are of course
used to identifying isomorphic structures in practice, but they generally do
so by \textquotedblleft abuse of notation\textquotedblright, or some other
informal device, knowing that the objects involved are not \textquotedblleft
really\textquotedblright\ identical. But in this new foundational scheme, such
structures can be formally identified, in the logical sense that every
property or construction involving one also applies to the other. \cite[p.
5]{ufp:hott}
\end{quotation}

Our purpose is rather more modest, to model the theory of paradigm-universals
$u_{S}$ and their projections $u_{T}$--that is analogous to working with sets
and subsets, e.g., in a Boolean algebra of subsets. That is all we will need
to show that probability theory can be developed using paradigms $u_{S}$
instead of subset-events $S$, and to make the connection to quantum mechanics.

Another homotopy example is `\textit{the} path going once (clockwise) around
the hole' in an annulus $A$ (disk with one hole), an abstract entity $1\in
\pi_{0}\left(  A\right)  \cong%
\mathbb{Z}
$:%

\begin{center}
\includegraphics[
height=1.2073in,
width=0.9729in
]%
{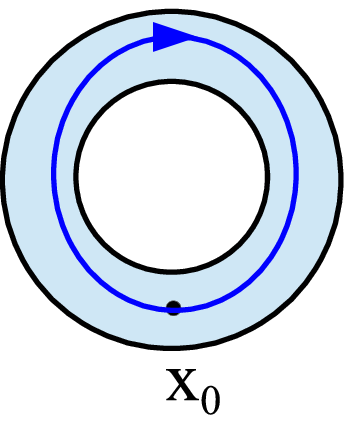}%
\end{center}

\begin{center}
Figure 2: `\textit{the} path going once (clockwise) around the hole'
\end{center}

\noindent Note that `\textit{the} path going once (clockwise) around the hole'
has the paradigmatic property of "going once (clockwise) around the hole" but
is \textit{not} one of the particular (coordinatized) paths that constitute
the equivalence class of coordinatized once-around paths deformable into one another.

In a similar manner, we can view other common $\#2$ abstractions such as:
`\textit{the} cardinal number $5$' that captures what is common to the
isomorphism class of all five-element sets; `\textit{the} number $1$
$\operatorname{mod}\left(  n\right)  $' that captures what is common within
the equivalence class $\left\{  ...,-2n+1,-n+1,1,n+1,2n+1,...\right\}  $ of
integers; `\textit{the} circle' or `\textit{the} equilateral triangle'--and so forth.

Category theory helped to motivate homotopy type theory for good reason.
Category theory has no notion of identity between objects, only isomorphism as
`equivalence' between objects. Therefore category theory can be seen as a
theory of \textit{abstract} $\#2$ objects ("up to isomorphism"), e.g.,
abstract sets, groups, spaces, etc.

\section{The Connection to Interpreting Symmetry Operations}

The difference between the $\#1$ abstract set and the $\#2$ abstract entity
can also be visually illustrated in a simple example of the symmetry operation
(defining an equivalence relation) of reflection on the $aA$-axis for a fully
definite isosceles triangles:%

\begin{center}
\includegraphics[
height=1.209in,
width=2.9482in
]%
{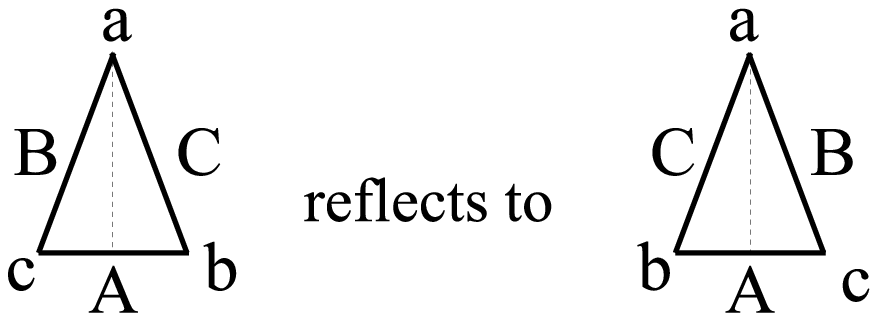}%
\end{center}

\begin{center}
Figure 3: Reflection on vertical axis symmetry operation.
\end{center}

\noindent Thus the equivalence class of reflective-symmetric figures in the
$\#1$ or classical interpretation is the set:%

\begin{center}
\includegraphics[
height=1.1606in,
width=2.4569in
]%
{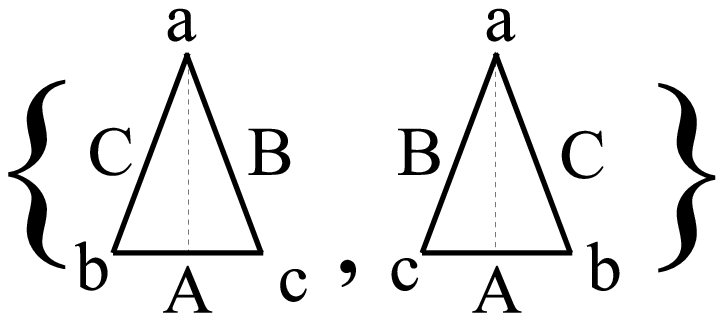}%
\end{center}

\begin{center}
Figure 4: The $\#1$ abstraction of equivalence class.
\end{center}

\noindent But under the $\#2$ or indefiniteness-abstraction(-quantum)
interpretation, the equivalence abstracts to the figure that is definite as to
what is the same and indefinite as to what is different between the definite
figures in the equivalence class:%

\begin{center}
\includegraphics[
height=1.1805in,
width=2.6403in
]%
{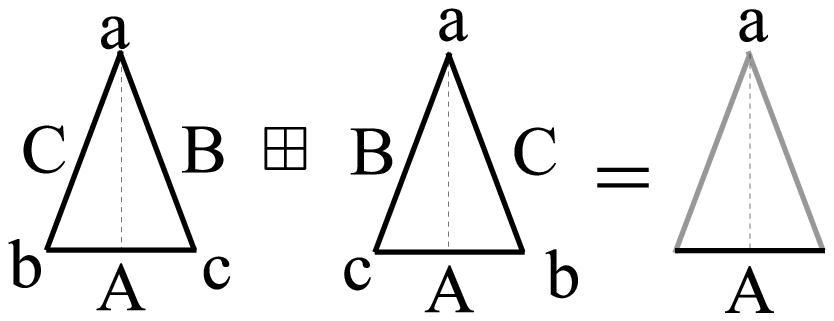}%
\end{center}

\begin{center}
Figure 5: The $\#2$ abstraction of indefinite entity.
\end{center}

\noindent Note that the symmetry operation on the indefinite figure is the
\textit{identity}. As noted in the discussion of homotopy type theory, the
movement from the $\#1$ equivalence class $S$ to the $\#2$ abstract-indefinite
entity $u_{S}$ replaces equivalence with identity. That is because the
symmetry operation goes from one element in an equivalence class $S$ to
another element in $S$ that differs in some definite aspects, but those are
precisely the aspects that are removed in the indefinite-abstract $u_{S}$--so
the symmetry just takes $u_{S}$ to itself.

Since we are later going to relate the $\#2$ entities to the indefinite states
of quantum mechanics, the example suggests that while classically a symmetry
operation is invariant on an equivalence class $S$ (i.e., takes one definite
element in the equivalence class $S$ to another definite element in $S$), in
the $\#2$ quantum case, the symmetry operation on the indefinite entity
$u_{S}$ is the \textit{identity}.

This is illustrated in the transition from the classical Maxwell-Boltzmann
statistics to the quantum Bose-Einstein statistics. Suppose we have two
particles of the same type which are classically indistinguishable so,
following Weyl, we distinguish them as Mike and Ike. If each of the two
particles could be in states $A$, $B$, or $C$, then the set of possible states
is the set of nine ordered pairs $\left\{  A,B,C\right\}  \times\left\{
A,B,C\right\}  $. Applying the symmetry operation of permuting Mike and Ike,
we have six equivalence classes.

\begin{center}%
\begin{tabular}
[c]{|c|c|}\hline
Equivalence classes under permutation & M-B\\\hline\hline
$\left\{  \left(  A,B\right)  ,\left(  B,A\right)  \right\}  $ & $\frac{2}{9}%
$\\\hline
$\left\{  \left(  A,C\right)  ,\left(  C,A\right)  \right\}  $ & $\frac{2}{9}%
$\\\hline
$\left\{  \left(  B,C\right)  ,\left(  C,B\right)  \right\}  $ & $\frac{2}{9}%
$\\\hline
$\left\{  \left(  A,A\right)  \right\}  $ & $\frac{1}{9}$\\\hline
$\left\{  \left(  B,B\right)  \right\}  $ & $\frac{1}{9}$\\\hline
$\left\{  \left(  C,C\right)  \right\}  $ & $\frac{1}{9}$\\\hline
\end{tabular}

Table 3: Maxwell-Boltzmann distribution.
\end{center}

\noindent Since the primitive data are the ordered pairs, we assign the equal
probabilities of $\frac{1}{9}$ to each pair which results in the
Maxwell-Boltzmann distribution for the equivalence classes.

But in the quantum case, we don't have an equivalence class $S$ of distinct
ordered pairs like $\left\{  \left(  A,B\right)  ,\left(  B,A\right)
\right\}  $ under the symmetry; we have a single indefinite entity
$u_{\left\{  \left(  A,B\right)  ,\left(  B,A\right)  \right\}  }$ where the
symmetry operation is the \textit{identity}. Since there are now only six
primitive entities, we assign the equal probabilities of $\frac{1}{6}$ to each
entity and obtain the Bose-Einstein distribution.

\begin{center}%
\begin{tabular}
[c]{|c|c|}\hline
Six indefinite states & B-E\\\hline\hline
\multicolumn{1}{|c|}{$u_{\left\{  \left(  A,B\right)  ,\left(  B,A\right)
\right\}  }$} & $\frac{1}{6}$\\\hline
\multicolumn{1}{|c|}{$u_{\left\{  \left(  A,C\right)  ,\left(  C,A\right)
\right\}  }$} & $\frac{1}{6}$\\\hline
\multicolumn{1}{|c|}{$u_{\left\{  \left(  B,C\right)  ,\left(  C,B\right)
\right\}  }$} & $\frac{1}{6}$\\\hline
\multicolumn{1}{|c|}{$u_{\left\{  \left(  A,A\right)  \right\}  }$} &
$\frac{1}{6}$\\\hline
\multicolumn{1}{|c|}{$u_{\left\{  \left(  B,B\right)  \right\}  }$} &
$\frac{1}{6}$\\\hline
\multicolumn{1}{|c|}{$u_{\left\{  \left(  C,C\right)  \right\}  }$} &
$\frac{1}{6}$\\\hline
\end{tabular}

Table 4: Bose-Einstein distribution.
\end{center}

Ruling out repeated states (i.e., the Pauli exclusion principle), there are
only three primitive entities and that gives the Fermi-Dirac
distribution.\footnote{For more of this pedagogical model of QM using sets
(where the sets may be given the $\#2$ abstraction $u_{S}$ interpretation),
see \cite{ell:qm-sets}.}

\begin{center}%
\begin{tabular}
[c]{|c|c|}\hline
Three possible indefinite states & F-D\\\hline\hline
\multicolumn{1}{|c|}{$u_{\left\{  \left(  A,B\right)  ,\left(  B,A\right)
\right\}  }$} & $\frac{1}{3}$\\\hline
\multicolumn{1}{|c|}{$u_{\left\{  \left(  A,C\right)  ,\left(  C,A\right)
\right\}  }$} & $\frac{1}{3}$\\\hline
\multicolumn{1}{|c|}{$u_{\left\{  \left(  B,C\right)  ,\left(  C,B\right)
\right\}  }$} & $\frac{1}{3}$\\\hline
\end{tabular}

Table 5: Fermi-Dirac distribution.
\end{center}

\section{How to Model the \#1 and \#2 Abstracts}

There are simple but different models to distinguish the $\#1$ and $\#2$
interpretations for $S\subseteq U$ with a finite $U=\left\{  u_{1}%
,...,u_{n}\right\}  $ such as:%

\begin{center}
\includegraphics[
height=0.6304in,
width=3.25in
]%
{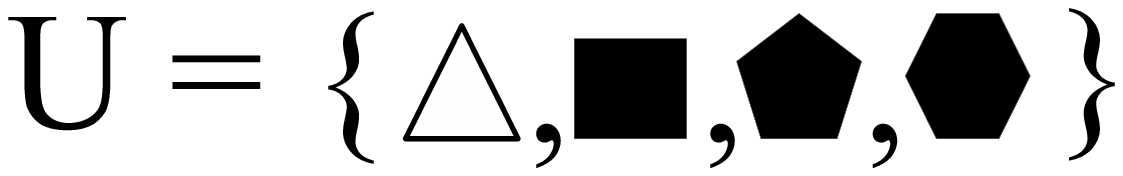}%
\end{center}

\begin{center}
Figure 6: Universe $U$ of figures
\end{center}

\noindent Ordinarily the set of solid figures $S=\left\{  u_{2},u_{3}%
,u_{4}\right\}  \subseteq\left\{  u_{1},u_{2},u_{3},u_{4}\right\}  =U$ would
be represented by a one-dimensional column vector $\left\vert S\right\rangle =%
\begin{bmatrix}
0\\
1\\
1\\
1
\end{bmatrix}
$, but by using a \textit{two}-dimensional matrix, we can \textit{represent}
the \textit{two} $\#1$ and $\#2$ versions of $S$ as two types of incidence matrices.

\begin{enumerate}
\item The $\#1$ (classical) representation of $S$ (i.e., set of $S$-things or
set of solid figures) is the diagonal matrix $\operatorname{In}\left(  \Delta
S\right)  $ that lays the column vector $\left\vert S\right\rangle $ along the
diagonal: $\operatorname{In}\left(  \Delta S\right)  =%
\begin{bmatrix}
0 & 0 & 0 & 0\\
0 & 1 & 0 & 0\\
0 & 0 & 1 & 0\\
0 & 0 & 0 & 1
\end{bmatrix}
=$ representation of set $S$ of distinct $S$-entities. $\operatorname{In}%
\left(  \Delta S\right)  $ is the \textit{incidence matrix} of the diagonal
$\Delta S\subseteq U\times U$ whose entries are the values of the
characteristic function $\chi_{\Delta S}\left(  u_{j},u_{k}\right)
=\delta_{jk}\chi_{S}\left(  u_{j}\right)  $.

\item The $\#2$ (quantum) representation of $S$ (i.e., \textit{the} $S$-thing)
is the matrix $\operatorname{In}\left(  S\times S\right)  $ that uses a $1$ in
the row $j$, column $k$ cell to mean $u_{j}$ and $u_{k}$ are both in $S$:
$\operatorname{In}\left(  S\times S\right)  =\left\vert S\right\rangle \left(
\left\vert S\right\rangle \right)  ^{t}=%
\begin{bmatrix}
0 & 0 & 0 & 0\\
0 & 1 & 1 & 1\\
0 & 1 & 1 & 1\\
0 & 1 & 1 & 1
\end{bmatrix}
=$ representation of one indistinct $S$-thing, `\textit{the} solid figure'
$u_{S}=u_{2}\boxplus u_{3}\boxplus u_{4}$. $\operatorname{In}\left(  S\times
S\right)  $ is the \textit{incidence matrix} of the product $S\times
S\subseteq U\times U$ (instead of the diagonal $\Delta S$) with the entries
$\chi_{S\times S}\left(  u_{j},u_{k}\right)  $.
\end{enumerate}

\noindent Note that for singletons $S=\left\{  u_{j}\right\}  $,
$\operatorname{In}\left(  \Delta S\right)  =$ $\operatorname{In}\left(
S\times S\right)  $ as expected, and for $\left\vert S\right\vert >1$,
$\operatorname{In}\left(  \Delta S\right)  \neq$ $\operatorname{In}\left(
S\times S\right)  $.

The two representations differ only in the off-diagonal entries. Think of the
off-diagonal $\operatorname{In}\left(  S\times S\right)  _{j,k}=1$'s as
equating, cohering, or `blobbing' together $u_{j}$ and $u_{k}$:

\begin{center}
$\operatorname{In}\left(  S\times S\right)  =%
\begin{bmatrix}
0 & 0 & 0 & 0\\
0 & 1 & 1 & 1\\
0 & 1 & 1 & 1\\
0 & 1 & 1 & 1
\end{bmatrix}
$ says

$%
\begin{bmatrix}
0 & 0 & 0 & 0\\
0 & 1 & {\small u}_{2}\thicksim{\small u}_{3} & {\small u}_{{\small 2}%
}\thicksim{\small u}_{{\small 4}}\\
0 & {\small u}_{{\small 3}}\thicksim{\small u}_{{\small 2}} & 1 &
{\small u}_{3}\thicksim{\small u}_{{\small 4}}\\
0 & {\small u}_{{\small 4}}\thicksim{\small u}_{{\small 2}} & {\small u}%
_{4}\thicksim{\small u}_{3} & 1
\end{bmatrix}
$.
\end{center}

We now can represent the blob-sum $\#2$ operation on entities: $u_{S}%
=\boxplus\left\{  u_{i}\in U|S\left(  u_{i}\right)  \right\}  $ as the
\textit{blob-sum} $\boxplus$ of the corresponding incidence matrices:

\begin{center}
$\operatorname{In}\left(  S\times S\right)  =\boxplus_{u_{i}\in S}$
$\operatorname{In}\left(  \left\{  u_{i}\right\}  \times\left\{
u_{i}\right\}  \right)  $
\end{center}

\noindent where the \textit{blob-sum} $\boxplus$ is defined for $S_{1}%
,S_{2}\subseteq U$ with $S=S_{1}\cup S_{2}$:

\begin{center}
$\operatorname{In}\left(  S_{1}\times S_{1}\right)  \boxplus\operatorname{In}%
\left(  S_{2}\times S_{2}\right)  :=$ $\operatorname{In}\left(  S\times
S\right)  =\operatorname{In}\left(  \left(  S_{1}\cup S_{2}\right)
\times\left(  S_{1}\cup S_{2}\right)  \right)  $

$=\operatorname{In}\left(  S_{1}\times S_{1}\cup S_{2}\times S_{2}\cup
S_{1}\times S_{2}\cup S_{2}\times S_{1}\right)  $

$=\operatorname{In}\left(  S_{1}\times S_{1}\right)  \vee\operatorname{In}%
\left(  S_{2}\times S_{2}\right)  \vee\operatorname{In}\left(  S_{1}\times
S_{2}\right)  \vee\operatorname{In}\left(  S_{2}\times S_{1}\right)  $.

Disjunction: $\operatorname{In}\left(  S_{1}\times S_{1}\right)
\vee\operatorname{In}\left(  S_{2}\times S_{2}\right)  \vee$ blobbing
cross-terms.\footnote{The disjunction of incidence matrices is the usual
entry-wise disjunction: $1\vee1=1\vee0=0\vee1=1$ and $0\vee0=0$, and similarly
for conjunction.}
\end{center}

For $S=\left\{  u_{2},u_{4}\right\}  $, the blob-sum $u_{S}=u_{2}\boxplus
u_{4}$ is represented by:

\begin{center}
$\operatorname{In}\left(  \left\{  u_{2}\right\}  \times\left\{
u_{2}\right\}  \right)  \boxplus\operatorname{In}\left(  \left\{
u_{4}\right\}  \times\left\{  u_{4}\right\}  \right)  =\operatorname{In}%
\left(  S\times S\right)  $
\end{center}

\noindent where the blob-sum operation $\boxplus$ means `blobbing-out' the
distinctions between entities in $S$ (given by the cross-terms in $\left\{
u_{2},u_{4}\right\}  \times\left\{  u_{2},u_{4}\right\}  $):

\begin{center}
$\operatorname{In}\left(  S\times S\right)  =\operatorname{In}\left(  \left\{
u_{2}\right\}  \times\left\{  u_{2}\right\}  \right)  \boxplus
\operatorname{In}\left(  \left\{  u_{4}\right\}  \times\left\{  u_{4}\right\}
\right)  $

$=%
\begin{bmatrix}
0 & 0 & 0 & 0\\
0 & 1 & 0 & 0\\
0 & 0 & 0 & 0\\
0 & 0 & 0 & 0
\end{bmatrix}
\boxplus%
\begin{bmatrix}
0 & 0 & 0 & 0\\
0 & 0 & 0 & 0\\
0 & 0 & 0 & 0\\
0 & 0 & 0 & 1
\end{bmatrix}
$

$=\operatorname{In}\left(  \left\{  u_{2},u_{4}\right\}  \times\left\{
u_{2},u_{4}\right\}  \right)  $

$=\operatorname{In}\left(  \left\{  u_{2}\right\}  \times\left\{
u_{2}\right\}  \right)  \vee\operatorname{In}\left(  \left\{  u_{4}\right\}
\times\left\{  u_{4}\right\}  \right)  \vee\operatorname{In}\left(  \left\{
u_{2}\right\}  \times\left\{  u_{4}\right\}  \right)  \vee\operatorname{In}%
\left(  \left\{  u_{4}\right\}  \times\left\{  u_{2}\right\}  \right)  $

$=%
\begin{bmatrix}
0 & 0 & 0 & 0\\
0 & 1 & 0 & 1\\
0 & 0 & 0 & 0\\
0 & 1 & 0 & 1
\end{bmatrix}
$.
\end{center}

Due to the development of Boolean subset logic and set theory, we are
perfectly comfortable with considering the $\#1$ abstractions of sets $S$ of
even concrete ur-elements like the set of entities on a table. The
representatives $\operatorname{In}\left(  \Delta S\right)  $ trivially form a
BA isomorphic to the BA of subsets $\wp\left(  U\right)  .$

To better understand abstraction in mathematics and indefinite states in QM,
we should become as comfortable with paradigms $u_{S}$ as with sets $S$. The
paradigms $u_{S}$ for $S\in\wp\left(  U\right)  $ form a Boolean algebra
isomorphic to $\wp\left(  U\right)  $ under the mapping: for any Boolean
operation $S\#T$ for $S,T\in\wp\left(  U\right)  $, $u_{S}\#u_{T}$ is the
paradigm represented by $\operatorname{In}\left(  \left(  S\#T\right)
\times\left(  S\#T\right)  \right)  $.

\begin{itemize}
\item The union of subsets $S\cup T$ induces the operation on paradigms
represented by $\operatorname{In}\left(  \left(  S\cup T\right)  \times\left(
S\cup T\right)  \right)  =\operatorname{In}\left(  S\times S\right)
\boxplus\operatorname{In}\left(  T\times T\right)  $, so the \textit{union or
join of paradigms} is the blob-sum $u_{S\cup T}$ $=u_{S}\boxplus u_{T}$ (note
as expected, for $T\subseteq S$, $u_{S}\boxplus u_{T}=u_{S}$);

\item The \textit{intersection or meet of paradigms} $u_{S}\wedge
u_{T}=u_{S\cap T}$ is represented by $\operatorname{In}\left(  S\cap T\times
S\cap T\right)  =\operatorname{In}\left(  S\times S\right)  \wedge
\operatorname{In}\left(  T\times T\right)  $ (note as expected, for
$T\subseteq S$, $u_{S}\wedge u_{T}=u_{T}$);

\item The \textit{negation of a paradigm} $\lnot u_{S}=u_{S^{c}}$ is
represented by $\operatorname{In}\left(  S^{c}\times S^{c}\right)
=\boxplus\left\{  \operatorname{In}\left(  \left\{  u\right\}  \times\left\{
u\right\}  \right)  |u\notin S\right\}  $ (note as expected, $u_{S}\boxplus
u_{S^{c}}=u_{U}$).
\end{itemize}

\section{The Projection Operation: Making an indefinite entity more definite}

Now suppose we classify or partition \textit{all }the elements of $U$
according to an attribute such as the parity of the number of sides, where a
\textit{partition} is a set of disjoint subsets (blocks) of $U$ whose union is
all of $U$. Let $\pi$ be the partition of two \textit{blocks }$O=\left\{
Odd\right\}  =\left\{  u_{1},u_{3}\right\}  $ and\textit{ }$E=\left\{
Even\right\}  =\left\{  u_{2},u_{4}\right\}  $.

The equivalence relation defined by $\pi$ is $\operatorname*{indit}\left(
\pi\right)  =\left(  O\times O\right)  \cup\left(  E\times E\right)  $
\cite{ell:partitions} and the disjunction is:

\begin{center}
$\operatorname{In}\left(  O\times O\right)  \vee\operatorname{In}\left(
E\times E\right)  =\operatorname{In}\left(  \operatorname*{indit}\left(
\pi\right)  \right)  $

$%
\begin{bmatrix}
1 & 0 & 1 & 0\\
0 & 0 & 0 & 0\\
1 & 0 & 1 & 0\\
0 & 0 & 0 & 0
\end{bmatrix}
\vee%
\begin{bmatrix}
0 & 0 & 0 & 0\\
0 & 1 & 0 & 1\\
0 & 0 & 0 & 0\\
0 & 1 & 0 & 1
\end{bmatrix}
=%
\begin{bmatrix}
1 & 0 & 1 & 0\\
0 & 1 & 0 & 1\\
1 & 0 & 1 & 0\\
0 & 1 & 0 & 1
\end{bmatrix}
$.
\end{center}

The $\#1$ (classical) operation of intersecting the set of even-sided figures
with the set of solid figures to give the set of even-sided solid figures is
represented as the conjunction:

\begin{center}
$\operatorname{In}\left(  \Delta E\right)  \wedge\operatorname{In}\left(
\Delta S\right)  =%
\begin{bmatrix}
0 & 0 & 0 & 0\\
0 & 1 & 0 & 0\\
0 & 0 & 0 & 0\\
0 & 0 & 0 & 1
\end{bmatrix}
\wedge%
\begin{bmatrix}
0 & 0 & 0 & 0\\
0 & 1 & 0 & 0\\
0 & 0 & 1 & 0\\
0 & 0 & 0 & 1
\end{bmatrix}
=%
\begin{bmatrix}
0 & 0 & 0 & 0\\
0 & 1 & 0 & 0\\
0 & 0 & 0 & 0\\
0 & 0 & 0 & 1
\end{bmatrix}
$.
\end{center}

The $\#2$ (quantum) operation of `sharpening' or `rendering more definite'
`\textit{the} solid figure' $u_{S}$ to `\textit{the} even-sided solid figure'
$u_{\left\{  u_{2},u_{4}\right\}  }$, so $u_{\left\{  u_{2},u_{4}\right\}
}\lhd u_{S}$ (suggested reading: $u_{\left\{  u_{2},u_{4}\right\}  }$ is a
projection of $u_{S}$) is represented as:

\begin{center}
$\operatorname{In}\left(  E\times E\right)  \wedge\operatorname{In}\left(
S\times S\right)  =%
\begin{bmatrix}
0 & 0 & 0 & 0\\
0 & 1 & 0 & 1\\
0 & 0 & 0 & 0\\
0 & 1 & 0 & 1
\end{bmatrix}
\wedge%
\begin{bmatrix}
0 & 0 & 0 & 0\\
0 & 1 & 1 & 1\\
0 & 1 & 1 & 1\\
0 & 1 & 1 & 1
\end{bmatrix}
=%
\begin{bmatrix}
0 & 0 & 0 & 0\\
0 & 1 & 0 & 1\\
0 & 0 & 0 & 0\\
0 & 1 & 0 & 1
\end{bmatrix}
$.
\end{center}

But there is a better way to represent `sharpening' using matrix
multiplication instead of just the logical operation $\wedge$ on matrices, and
it foreshadows the measurement operation in QM. The matrix $\operatorname{In}%
\left(  \Delta E\right)  =P_{E}$ is a projection matrix, i.e., the diagonal
matrix with diagonal entries $\chi_{E}\left(  u_{i}\right)  $ so
$P_{E}\left\vert S\right\rangle =\left\vert E\cap S\right\rangle $. Then the
result of the projection-sharpening can be represented as:

\begin{center}
$\left\vert E\cap S\right\rangle \left(  \left\vert E\cap S\right\rangle
\right)  ^{t}=P_{E}\left\vert S\right\rangle \left(  P_{E}\left\vert
S\right\rangle \right)  ^{t}=P_{E}\left\vert S\right\rangle \left(  \left\vert
S\right\rangle \right)  ^{t}P_{E}$

$=P_{E}\operatorname{In}\left(  S\times S\right)  P_{E}=\operatorname{In}%
\left(  E\times E\right)  \wedge\operatorname{In}\left(  S\times S\right)  $.
\end{center}

\noindent Under the $\#2$ interpretation, the parity-sharpening,
parity-differentiation, or \textit{parity-measurement} of `\textit{the} solid
figure' by both parities is represented as:

\begin{center}
$\operatorname{In}\left(  \operatorname*{indit}\left(  \pi\right)  \right)
\wedge$ $\operatorname{In}\left(  S\times S\right)  =P_{O}\operatorname{In}%
\left(  S\times S\right)  P_{O}+P_{E}\operatorname{In}\left(  S\times
S\right)  P_{E}$

$=%
\begin{bmatrix}
1 & 0 & 1 & 0\\
0 & 1 & 0 & 1\\
1 & 0 & 1 & 0\\
0 & 1 & 0 & 1
\end{bmatrix}
\wedge%
\begin{bmatrix}
0 & 0 & 0 & 0\\
0 & 1 & 1 & 1\\
0 & 1 & 1 & 1\\
0 & 1 & 1 & 1
\end{bmatrix}
=%
\begin{bmatrix}
0 & 0 & 0 & 0\\
0 & 1 & 0 & 1\\
0 & 0 & 1 & 0\\
0 & 1 & 0 & 1
\end{bmatrix}
$.
\end{center}

\noindent The results are `\textit{the} even-sided solid figure' $u_{\left\{
u_{2},u_{4}\right\}  }$ and `\textit{the} odd-sided solid figure' $u_{\left\{
u_{3}\right\}  }=u_{3}$. The important thing to notice is the action on the
off-diagonal elements where the action $1\rightsquigarrow0$ in the $j,k$-entry
means that $u_{j}$ and $u_{k}$ have been deblobbed, decohered, distinguished,
or differentiated--in this case by parity:

\begin{center}
$\operatorname{In}\left(  S\times S\right)  \leadsto$ $\operatorname{In}%
\left(  \operatorname*{indit}\left(  \pi\right)  \right)  \wedge$
$\operatorname{In}\left(  S\times S\right)  $

$=P_{O}\operatorname{In}\left(  S\times S\right)  P_{O}+P_{E}\operatorname{In}%
\left(  S\times S\right)  P_{E}$

$%
\begin{bmatrix}
0 & 0 & 0 & 0\\
0 & 1 & 1\overset{{\tiny deblob}}{\leadsto}0 & 1\\
0 & 1\overset{{\tiny deblob}}{\leadsto}0 & 1 & 1\overset{{\tiny deblob}%
}{\leadsto}0\\
0 & 1 & 1\overset{{\tiny deblob}}{\leadsto}0 & 1
\end{bmatrix}
$.
\end{center}

We could also classify the figures as to having $4$ or fewer sides ("few
sides") or not ("many sides") so that partition is $\sigma=\left\{  \left\{
u_{1},u_{2}\right\}  ,\left\{  u_{3},u_{4}\right\}  \right\}  $ which is
represented by:

\begin{center}
$\operatorname{In}\left(  \operatorname*{indit}\left(  \sigma\right)  \right)
=%
\begin{bmatrix}
1 & 1 & 0 & 0\\
1 & 1 & 0 & 0\\
0 & 0 & 1 & 1\\
0 & 0 & 1 & 1
\end{bmatrix}
$ and

$\operatorname{In}\left(  \operatorname*{indit}\left(  \sigma\right)  \right)
\wedge\left(  \operatorname{In}\left(  \operatorname*{indit}\left(
\pi\right)  \right)  \wedge\operatorname{In}\left(  S\times S\right)  \right)
=%
\begin{bmatrix}
0 & 0 & 0 & 0\\
0 & 1 & 0 & 0\\
0 & 0 & 1 & 0\\
0 & 0 & 0 & 1
\end{bmatrix}
=$ $\operatorname{In}\left(  \Delta S\right)  $.
\end{center}

Thus parity and few-or-many-sides properties suffice to classify the solid
figures uniquely and thus to yield the representation $\operatorname{In}%
\left(  \Delta S\right)  $ of the distinct elements of $S=\left\{  u_{2}%
,u_{3},u_{4}\right\}  $. Thus making all the distinctions (i.e., decohering
the entities that cohered together in $u_{S}$) takes $\operatorname{In}\left(
S\times S\right)  \leadsto$ $\operatorname{In}\left(  \Delta S\right)  $.

In QM jargon, the parity and few-or-many-sides attributes constitute a
"complete set of commuting operators" (CSCO) so that measurement of
`\textit{the} solid figure' by those observables will take `\textit{the} solid
figure,' to the separate eigen-solid-figures: `\textit{the} few- and
even-sided solid figure' (the square $u_{2}$), `\textit{the} many- and
odd-sided solid figure' (the pentagon $u_{3}$), and `\textit{the} many- and
even-sided solid figure' (the hexagon $u_{4}$).

\section{From Incidence to Density Matrices}

The incidence matrices $\operatorname{In}\left(  \Delta S\right)  $ and
$\operatorname{In}\left(  S\times S\right)  $ can be turned into
\textit{density matrices} by dividing through by their trace:

\begin{center}
$\rho\left(  \Delta S\right)  =\frac{1}{\operatorname*{tr}\left[
\operatorname{In}\left(  \Delta S\right)  \right]  }$ $\operatorname{In}%
\left(  \Delta S\right)  $ and $\rho\left(  S\right)  =\frac{1}%
{\operatorname*{tr}\left[  \operatorname{In}\left(  S\times S\right)  \right]
}$ $\operatorname{In}\left(  S\times S\right)  $.
\end{center}

\noindent In terms of probabilities, this means treating the outcomes in $S$
as being equiprobable with probability $\frac{1}{\left\vert S\right\vert }$.
But now we have the $\#1$ and $\#2$ interpretations of the sample space for
finite discrete probability theory.

\begin{enumerate}
\item The $\#1$ (classical) interpretation, represented by $\rho\left(  \Delta
S\right)  $, is the classical version with $S$ as the sample space of
outcomes. For instance, the $6\times6$ diagonal matrix with diagonal entries
$\frac{1}{6}$ is "the statistical mixture describing the state of a classical
dice [die] before the outcome of the throw" \cite[p. 176]{auletta:qm};

\item The $\#2$ (quantum) interpretation replaces the "sample space" with the
one indefinite `\textit{the} sample outcome' $u_{S}$ represented by
$\rho\left(  S\right)  $ (like `\textit{the} outcome of throwing a die') and,
in a trial, the indefinite outcome $u_{S}$ `sharpens to' or becomes a definite
outcome $u_{i}\in S$ with probability $\frac{1}{\left\vert S\right\vert }$.
\end{enumerate}

Let $f:U\rightarrow%
\mathbb{R}
$ be a real-valued random variable with distinct values $\phi_{i}$ for
$i=1,...,m$ and let $\pi=\left\{  B_{i}\right\}  _{i=1,...,m}$ where
$B_{i}=f^{-1}\left(  \phi_{i}\right)  $, be the partition of $U$ according to
the values. The classification of $\rho\left(  S\right)  $ according to the
different values is: $\operatorname{In}\left(  \operatorname*{indit}\left(
\pi\right)  \right)  \wedge\rho\left(  S\right)  $ which distinguishes the
elements of $S$ that have different $f$-values. If $P_{B_{i}}$ is the diagonal
(projection) matrix with diagonal elements $\left(  P_{B_{i}}\right)
_{jj}=\chi_{B_{i}}\left(  u_{j}\right)  $, then the probability of a trial
returning a $u_{j}$ with $f\left(  u_{j}\right)  =\phi_{i}$ is:

\begin{center}
$\Pr\left(  \phi_{i}|S\right)  =\operatorname*{tr}\left[  P_{B_{i}}\rho\left(
S\right)  \right]  $.
\end{center}

For instance, in the previous example, where $f:U\rightarrow%
\mathbb{R}
$ gives the parity partition $\pi$ with the two values $\phi_{odd}$ and
$\phi_{even}$, then:

\begin{center}
$P_{even}\rho\left(  S\right)  =%
\begin{bmatrix}
0 & 0 & 0 & 0\\
0 & 1 & 0 & 0\\
0 & 0 & 0 & 0\\
0 & 0 & 0 & 1
\end{bmatrix}%
\begin{bmatrix}
0 & 0 & 0 & 0\\
0 & \frac{1}{3} & \frac{1}{3} & \frac{1}{3}\\
0 & \frac{1}{3} & \frac{1}{3} & \frac{1}{3}\\
0 & \frac{1}{3} & \frac{1}{3} & \frac{1}{3}%
\end{bmatrix}
\allowbreak=\allowbreak%
\begin{bmatrix}
0 & 0 & 0 & 0\\
0 & \frac{1}{3} & \frac{1}{3} & \frac{1}{3}\\
0 & 0 & 0 & 0\\
0 & \frac{1}{3} & \frac{1}{3} & \frac{1}{3}%
\end{bmatrix}
$
\end{center}

\noindent so $\operatorname*{tr}\left[  P_{even}\rho\left(  S\right)  \right]
=\frac{2}{3}$ which is the conditional probability of getting `\textit{the}
even-sided solid figure' starting with `\textit{the} solid figure' in the
$\#2$ (quantum) interpretation. And under the $\#1$ (standard) interpretation,
$\Pr\left(  \phi_{even}|S\right)  =\operatorname*{tr}\left[  P_{even}%
\rho\left(  \Delta S\right)  \right]  =\frac{2}{3}$ which is the probability
of getting \textit{an} even-sided solid figure starting with the set of solid figures.

These two interpretations of finite discrete probability theory extend easily
to the case of point probabilities $p_{j}$ for $u_{j}\in U$, where:

\begin{enumerate}
\item $\left(  \rho\left(  \Delta S\right)  \right)  _{jj}=\chi_{S}\left(
u_{j}\right)  p_{j}/\Pr\left(  S\right)  $, so $\operatorname*{tr}\left[
P_{even}\rho\left(  \Delta S\right)  \right]  =$ probability of getting
\textit{an}\textbf{ }even-sided solid figure starting with the set of solid
figures, and

\item $\left(  \rho\left(  S\right)  \right)  _{j,k}=\chi_{S}\left(
u_{j}\right)  \chi_{S}\left(  u_{k}\right)  \sqrt{p_{j}p_{k}}/\Pr\left(
S\right)  $, so $\operatorname*{tr}\left[  P_{even}\rho\left(  S\right)
\right]  =$ probability of getting \textbf{`}\textit{the}\textbf{ }even-sided
solid figure' starting with `\textit{the} solid figure.'
\end{enumerate}

\noindent The whole of finite discrete probability theory can be developed in
this manner, \textit{mutatis mutandis}, for the $\#2$ interpretation paradigms.

\section{Density matrices in Quantum Mechanics}

The jump to quantum mechanics (QM) is to replace the binary digits like $0,1$
in incidence matrices or reals $\sqrt{p_{j}p_{k}}$ in `classical' density
matrices by complex numbers. Instead of the set $S$ represented by a column
$\left\vert S\right\rangle $ of $0,1$, we have a normalized column $\left\vert
\psi\right\rangle $ of complex numbers $\alpha_{j}$ whose absolute squares are
probabilities: $\left\vert \alpha_{j}\right\vert ^{2}=p_{j}$, e.g.,

\begin{center}
$\left\vert S\right\rangle =%
\begin{bmatrix}
0\\
1\\
1\\
1
\end{bmatrix}
\leadsto\left\vert \psi\right\rangle =%
\begin{bmatrix}
\alpha_{1}\\
\alpha_{2}\\
\alpha_{3}\\
\alpha_{4}%
\end{bmatrix}
$

where $\alpha_{1}=0$ and $\left\vert \alpha_{j}\right\vert ^{2}=p_{j}$ for
$j=2,3,4$.
\end{center}

\begin{enumerate}
\item The \textit{density matrix} $\rho\left(  \Delta\psi\right)  $ has the
absolute squares $\left\vert \alpha_{j}\right\vert ^{2}=p_{j}$ laid out along
the diagonal.

\item The \textit{density matrix} $\rho\left(  \psi\right)  $ has the
$j,k$-entry as the product of $\alpha_{j}$ and $\alpha_{k}^{\ast}$ (complex
conjugate of $\alpha_{k}$), where $p_{j}=\alpha_{j}^{\ast}\alpha
_{j}=\left\vert \alpha_{j}\right\vert ^{2}$.
\end{enumerate}

Thus:

\begin{center}
$\rho\left(  \Delta\psi\right)  =%
\begin{bmatrix}
0 & 0 & 0 & 0\\
0 & p_{2} & 0 & 0\\
0 & 0 & p_{3} & 0\\
0 & 0 & 0 & p_{4}%
\end{bmatrix}
$ and $\rho\left(  \psi\right)  =%
\begin{bmatrix}
0 & 0 & 0 & 0\\
0 & p_{2} & \alpha_{2}\alpha_{3}^{\ast} & \alpha_{2}\alpha_{4}^{\ast}\\
0 & \alpha_{3}\alpha_{2}^{\ast} & p_{3} & \alpha_{3}\alpha_{4}^{\ast}\\
0 & \alpha_{4}\alpha_{2}^{\ast} & \alpha_{4}\alpha_{3}^{\ast} & p_{4}%
\end{bmatrix}
$.
\end{center}

\begin{quotation}
\noindent\lbrack The] off-diagonal terms of a density matrix...are often
called \textit{quantum coherences} because they are responsible for the
interference effects typical of quantum mechanics that are absent in classical
dynamics. \cite[p. 177]{auletta:qm}
\end{quotation}

The classifying or measuring operation $\operatorname{In}\left(
\operatorname*{indit}\left(  \pi\right)  \right)  \wedge\rho\left(
\psi\right)  $ could still be defined taking the minimum of corresponding
entries in absolute value, but in QM it is defined as the \textit{L\"{u}ders
mixture operation \cite[p. 279]{auletta:qm}}. If $\pi=\left\{  B_{1}%
,...,B_{m}\right\}  $ is a partition according to the eigenvalues $\phi
_{1},...,\phi_{m}$ on $U=\left\{  u_{1},...,u_{n}\right\}  $ (where $U$ is an
orthonormal basis set for the observable being measured), let $P_{B_{i}}$ be
the diagonal (projection) matrix with diagonal entries $\left(  P_{B_{i}%
}\right)  _{jj}=\chi_{B_{i}}\left(  u_{j}\right)  $. Then $\operatorname{In}%
\left(  \operatorname*{indit}\left(  \pi\right)  \right)  \wedge\rho\left(
\psi\right)  $ is obtained as:

\begin{center}
$\sum_{B_{i}\in\pi}P_{B_{i}}\rho\left(  \psi\right)  P_{B_{i}}$

The L\"{u}ders mixture.
\end{center}

\noindent The probability of getting the result $\phi_{i}$ is:

\begin{center}
$\Pr\left(  \phi_{i}|\psi\right)  =\operatorname*{tr}\left[  P_{B_{i}}%
\rho\left(  \psi\right)  \right]  $.
\end{center}

\section{A Pop Science Interlude}

The popular science version of the simplest case is Schr\"{o}dinger's cat.%

\begin{center}
\includegraphics[
height=2.1101in,
width=1.7218in
]%
{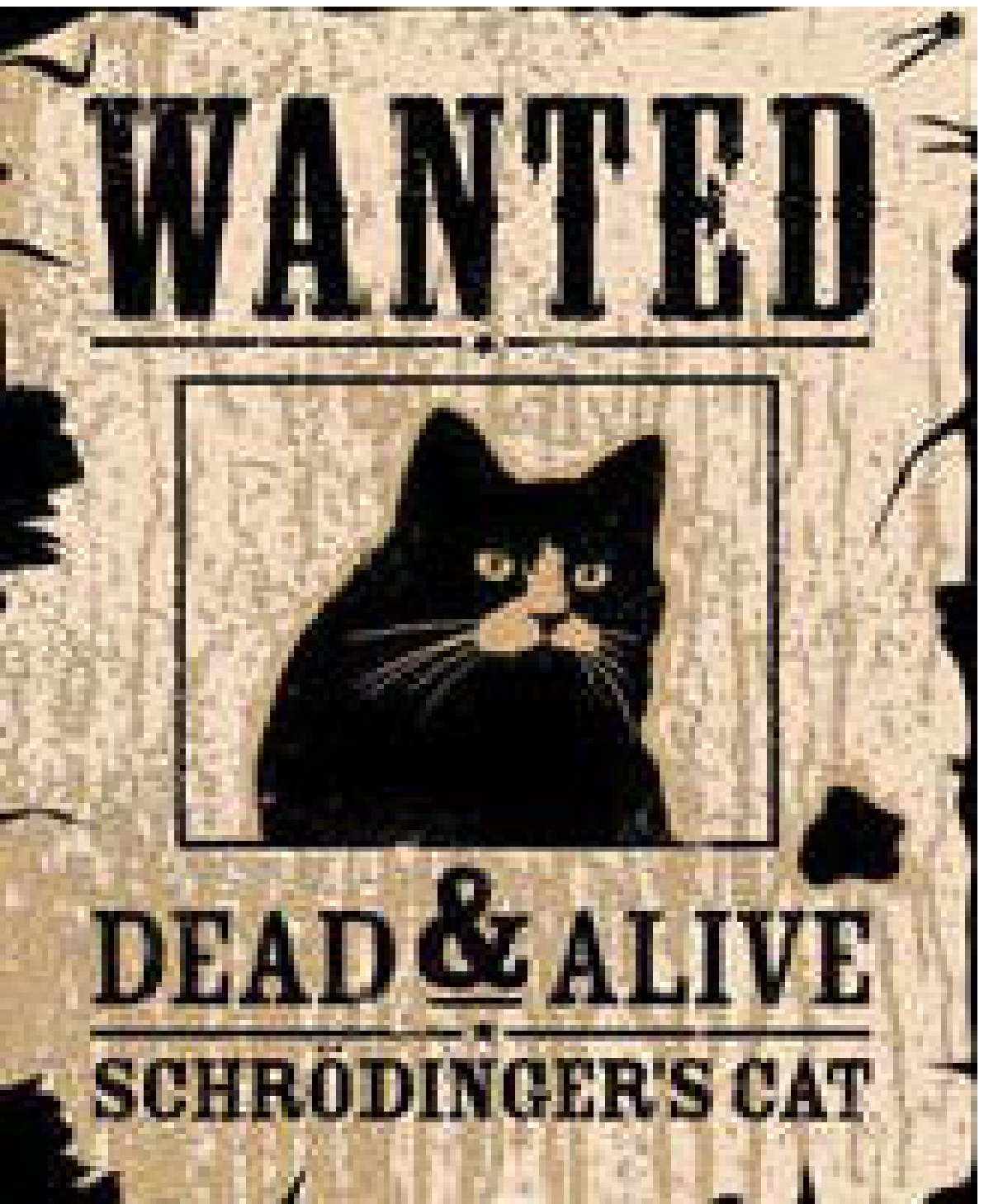}%
\end{center}

\begin{center}
Figure 7: Usual "And" version of Schr\"{o}dinger's cat.
\end{center}

\noindent This version of Schr\"{o}dinger's cat as being "Dead \& Alive" is
like the usual mis-interpretation of the unobserved particle as going through
"Slit 1 \& Slit 2" in the double slit experiment. But the cat is not
definitely alive \textbf{and} definitely dead at the same time. The quantum
version is that the cat is indefinite between those two definite
possibilities; it's in cat-limbo.

\begin{center}
Schr\"{o}dinger's cat = dead-cat $\boxplus$ live-cat.
\end{center}

It would be more accurate to say "Dead or Alive--but neither definitely," a
"disjunctive fact" \cite{stairs:disjfacts}.%

\begin{center}
\includegraphics[
height=1.8689in,
width=1.9424in
]%
{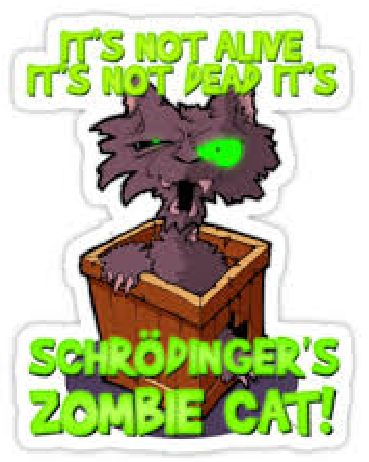}%
\end{center}

\begin{center}
Figure 8: The disjunctive cat.
\end{center}

Technically the state vector is:%

\begin{center}
\includegraphics[
height=1.1978in,
width=3.4307in
]%
{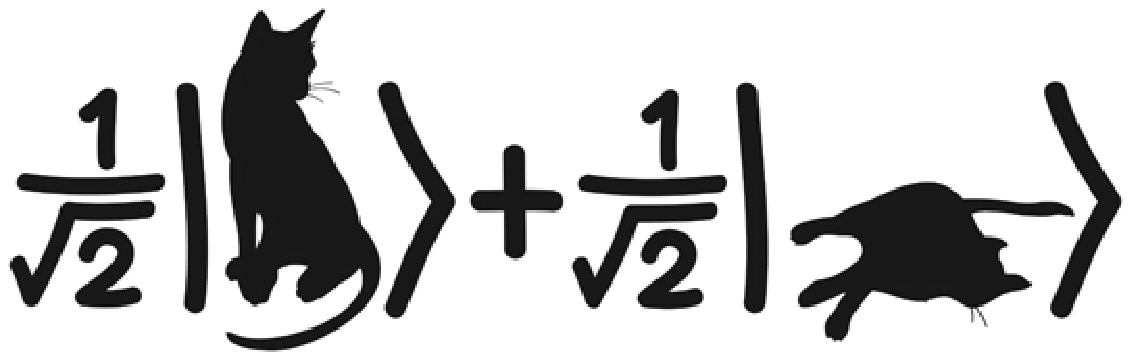}%
\end{center}

\begin{center}
Figure 9: Schr\"{o}dinger's cat state vector.
\end{center}

Using density matrices, we would represent Schr\"{o}dinger's cat as being in
the state:

\begin{center}
$\rho\left(  cat\right)  =%
\begin{bmatrix}
\frac{1}{2}live & \frac{1}{2}\\
\frac{1}{2} & \frac{1}{2}dead
\end{bmatrix}
$.
\end{center}

\section{Simplest Quantum Example}

Consider a system with two spin-observable $\sigma$ eigenstates$\ \left\vert
\uparrow\right\rangle $ and $\left\vert \downarrow\right\rangle $ (like
electron spin up or down along the $z$-axis) where the given normalized
superposition state is $\left\vert \psi\right\rangle =\frac{1}{\sqrt{2}%
}\left\vert \uparrow\right\rangle +\frac{1}{\sqrt{2}}\left\vert \downarrow
\right\rangle =%
\begin{bmatrix}
\alpha_{\uparrow}\\
\alpha_{\downarrow}%
\end{bmatrix}
=%
\begin{bmatrix}
\frac{1}{\sqrt{2}}\\
\frac{1}{\sqrt{2}}%
\end{bmatrix}
$ so the density matrix is $\rho\left(  \psi\right)  =%
\begin{bmatrix}
p_{\uparrow} & \alpha_{\uparrow}\alpha_{\downarrow}^{\ast}\\
\alpha_{\downarrow}\alpha_{\uparrow}^{\ast} & p_{\downarrow}%
\end{bmatrix}
=%
\begin{bmatrix}
\frac{1}{2} & \frac{1}{2}\\
\frac{1}{2} & \frac{1}{2}%
\end{bmatrix}
$ where $p_{\uparrow}=\alpha_{\uparrow}\alpha_{\uparrow}^{\ast}$ and
$p_{\downarrow}=\alpha_{\downarrow}\alpha_{\downarrow}^{\ast}$. The
measurement in that spin-observable $\sigma$ goes from $\rho\left(
\psi\right)  $ to

\begin{center}
$\operatorname{In}\left(  \operatorname*{indit}\left(  \sigma\right)  \right)
\wedge\rho\left(  \psi\right)  =%
\begin{bmatrix}
1 & 0\\
0 & 1
\end{bmatrix}
\wedge%
\begin{bmatrix}
p_{\uparrow} & \alpha_{\uparrow}\alpha_{\downarrow}^{\ast}\\
\alpha_{\downarrow}\alpha_{\uparrow}^{\ast} & p_{\downarrow}%
\end{bmatrix}
=%
\begin{bmatrix}
p_{\uparrow} & 0\\
0 & p_{\downarrow}%
\end{bmatrix}
=$ $%
\begin{bmatrix}
\frac{1}{2} & 0\\
0 & \frac{1}{2}%
\end{bmatrix}
=\rho\left(  \Delta\psi\right)  $.
\end{center}

Or using the L\"{u}ders mixture operation:

\begin{center}
$P_{\uparrow}\rho\left(  \psi\right)  P_{\uparrow}+P_{\downarrow}\rho\left(
\psi\right)  P_{\downarrow}$

$=%
\begin{bmatrix}
1 & 0\\
0 & 0
\end{bmatrix}%
\begin{bmatrix}
p_{\uparrow} & \alpha_{\uparrow}\alpha_{\downarrow}^{\ast}\\
\alpha_{\downarrow}\alpha_{\uparrow}^{\ast} & p_{\downarrow}%
\end{bmatrix}%
\begin{bmatrix}
1 & 0\\
0 & 0
\end{bmatrix}
+%
\begin{bmatrix}
0 & 0\\
0 & 1
\end{bmatrix}%
\begin{bmatrix}
p_{\uparrow} & \alpha_{\uparrow}\alpha_{\downarrow}^{\ast}\\
\alpha_{\downarrow}\alpha_{\uparrow}^{\ast} & p_{\downarrow}%
\end{bmatrix}%
\begin{bmatrix}
0 & 0\\
0 & 1
\end{bmatrix}
$

$=%
\begin{bmatrix}
p_{\uparrow} & 0\\
0 & p_{\downarrow}%
\end{bmatrix}
=$ $%
\begin{bmatrix}
\frac{1}{2} & 0\\
0 & \frac{1}{2}%
\end{bmatrix}
=\rho\left(  \Delta\psi\right)  $.
\end{center}

The two versions of $S=U$ give us two versions of finite discrete probability
theory where: $\#1)$ $U$ is the sample space or $\#2)$ $u_{U}$ is \textit{the}
sample outcome.

\begin{enumerate}
\item The $\#1$ classical version is the usual version which in this case is
like flipping a fair coin and getting head or tails with equal probability.
\end{enumerate}

%

\begin{center}
\includegraphics[
height=1.1234in,
width=3.5829in
]%
{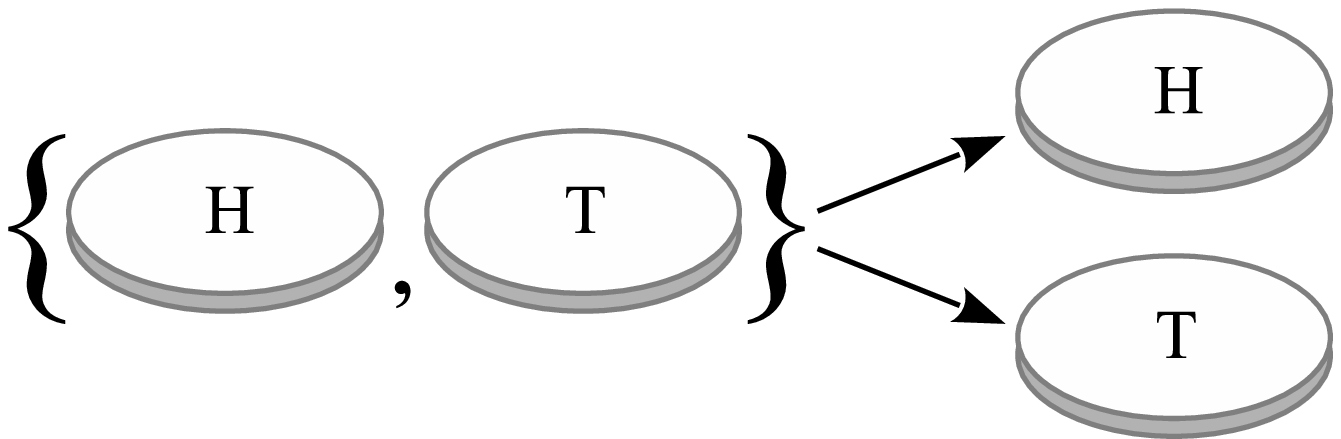}%
\end{center}

\begin{center}
Figure 10: Outcome set for classical coin-flipping trial.
\end{center}

\begin{enumerate}
\item[2.] The $\#2$ quantum version starts with the indefinite entity
$u_{U}=\boxplus\left\{  u_{i}\in U\right\}  $, `\textit{the} (indefinite)
outcome', and a trial renders it into one of the definite outcomes $u_{i}$
with some probability $p_{i}$ so that $u_{U}$ could be represented by the
density matrix $\rho\left(  U\right)  $ where $\left(  \rho\left(  U\right)
\right)  _{jk}=\sqrt{p_{j}p_{k}}$. In this case, this is like a coin
$u_{\left\{  H,T\right\}  }$ with the difference between heads or tails
rendered indefinite or blobbed out, and the trial results in it sharpening to
definitely heads or definitely tails with equal probability.
\end{enumerate}

%

\begin{center}
\includegraphics[
height=1.2704in,
width=3.0139in
]%
{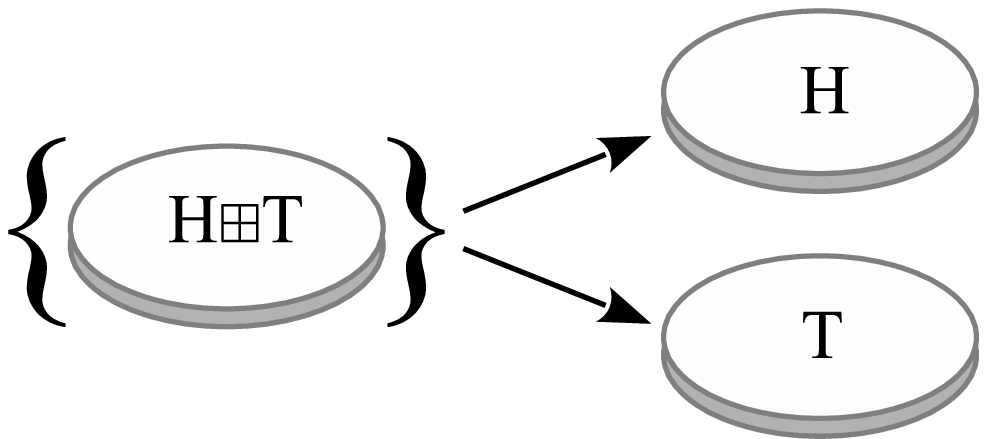}%
\end{center}

\begin{center}
Figure 11: `\textit{the} outcome state' for quantum coin-flipping trial.
\end{center}

Experimentally, it is not possible to distinguish between the $\#1$ and $\#2$
versions \textit{by }$\sigma$\textit{-measurements}. But in QM the two states
$\rho\left(  \Delta\psi\right)  $ and $\rho\left(  \psi\right)  $ \textit{can}
be distinguished by measuring other observables like spin along a different
axis \cite[p. 176]{auletta:qm}. Thus we \textit{know} in QM which version is
the superposition (pure) state $\left\vert \psi\right\rangle =%
\begin{bmatrix}
\alpha_{\uparrow}\\
\alpha_{\downarrow}%
\end{bmatrix}
$; it is \textit{the }$\#2$\textit{\ blob-state }$\rho\left(  \psi\right)  $.

\section{Conclusions}

Quantum mechanics texts usually mention several interpretations such as the
Copenhagen, many-worlds, or hidden-variables interpretations. Now that we have
established a bridge from abstraction in mathematics to indefinite states in
QM, we may (for fun) cross the bridge in the opposite direction. For instance,
in the many-worlds (or many-minds) interpretation, $1\in\pi_{0}\left(
A\right)  \cong%
\mathbb{Z}
$ would refer to a different specific coordinatized "once clockwise around the
hole" path in each different world (or mind).

Shimony, however, suggests the \textit{Literal or Objective Indefiniteness
Interpretation}--which we have seen is suggested by the mathematics itself.

\begin{quotation}
\noindent But the mathematical formalism ... suggests a philosophical
interpretation of quantum mechanics which I shall call "the Literal
Interpretation." ...This is the interpretation resulting from taking the
formalism of quantum mechanics literally, as giving a representation of
physical properties themselves, rather than of human knowledge of them, and by
taking this representation to be complete. \cite[pp. 6-7]{shim:vienna}
\end{quotation}

We have approached QM by starting with the logical situation of a universe $U$
of distinct entities. Given a property $S\left(  x\right)  $ on $U$, we can
associate with it:

\begin{enumerate}
\item the $\#1$ abstract object $S=\left\{  u_{i}\in U|S\left(  u_{i}\right)
\right\}  $, the set of $S\left(  x\right)  $-entities, or

\item the $\#2$ abstract object $u_{S}=\boxplus\left\{  u_{i}\in U|S\left(
u_{i}\right)  \right\}  $ which is the abstract entity expressing the
properties common to the $S\left(  x\right)  $-entities but "abstracting away
from," "rendering indefinite," "cohering together," or "blobbing out" the
differences between those entities.
\end{enumerate}

We argued that the mathematical formalisms of incidence matrices and then
density matrices can be used to formalize the two representations:

\begin{enumerate}
\item $\#1$ representation as $\operatorname{In}\left(  \Delta S\right)  $ or
$\rho\left(  \Delta\psi\right)  $; and

\item $\#2$ representation as $\operatorname{In}\left(  S\times S\right)  $ or
$\rho\left(  \psi\right)  $.
\end{enumerate}

\noindent This dove-tailed precisely into usual density-matrix treatment in QM
of quantum states $\left\vert \psi\right\rangle $ as $\rho\left(  \psi\right)
$ which, as suggested by Shimony, can be interpreted as \textit{objectively}
indefinite states.

Yet since the ancient Greeks, we have the $\#2$ Platonic notion of the
abstract paradigm-universal `\textit{the} $S$-entity', definite on what is
common to the members of a set $S$ and indefinite on where they differ, so the
connection that may help to better understand quantum mechanics is:

\begin{center}
The paradigm $u_{S}$, `\textit{the} $S$-entity' represented by
$\operatorname{In}\left(  S\times S\right)  $ $\iff$ the superposition state
$\psi$ represented by the density matrix $\rho\left(  \psi\right)  $.
\end{center}

\noindent This recalls Whitehead's quip that Western philosophy is "a series
of footnotes to Plato." \cite[p. 39]{whitehead:pandr}

\end{document}